\documentclass[EPJ,twocolumn]{webofc}
\usepackage[varg]{txfonts}   
\wocname{Atmohead Conference series}
\woctitle{Proceedings of the 2nd Atmohead Conference, May 19-21, 2014. Padova (Italy) }
\begin{document}
\title{The Atmospheric Monitoring System of the JEM-EUSO Space Mission}

\author{M. D. Rodr\'iguez Fr\'ias\inst{1,2} \and
  S. Toscano\inst{2} \and
  E. Bozzo\inst{2}\and
  L. del Peral\inst{1,2}\fnsep\thanks{\email{luis.delperal@gmail.com}} \and
  A. Neronov\inst{2}\and
S. Wada\inst{3} \ for the JEM-EUSO Collaboration.
}

\institute{SPace \& AStroparticle (SPAS) Group, UAH, Madrid, Spain.
  \and
  ISDC, Astronomy Dept. University of Geneva, Versoix, Switzerland.
  \and
  RIKEN Advanced Science Institute, Japan.
}

\abstract{%
 An Atmospheric Monitoring System (AMS) is a mandatory and key device of a space-based mission which
aims to detect Ultra-High Energy Cosmic Rays (UHECR) and Extremely-High Energy Cosmic Rays (EHECR)
from Space. JEM-EUSO has a dedicated atmospheric monitoring system that plays a fundamental role in our
understanding of the atmospheric conditions in the Field of View (FoV) of the telescope. Our AMS consists of
a very challenging space infrared camera and a LIDAR device, that are being fully designed with space
qualification to fulfil the scientific requirements of this space mission. The AMS will provide information of
the cloud cover in the FoV of JEM-EUSO, as well as measurements of the cloud top altitudes with an accuracy of 500 m and the optical depth profile of the atmosphere transmittance in the direction of each air shower
with an accuracy of 0.15 degree and a resolution of 500 m. This will ensure that the energy of the primary
UHECR and the depth of maximum development of the EAS ( Extensive Air Shower) are measured with an
accuracy better than 30\% primary energy and 120 $g/cm^2$ depth of maximum development for EAS occurring
either in clear sky or with the EAS depth of maximum development above optically thick cloud layers.
Moreover a very novel radiometric retrieval technique considering the LIDAR shots as calibration points, that seems to be the most promising retrieval algorithm is under development to infer the
Cloud Top Height (CTH) of all kind of clouds, thick and thin clouds in the FoV of the JEM-EUSO space telescope.
}
\maketitle

\section{Introduction}
\label{sec-intro}

Cosmic Ray Physics is one of the fundamental key issues and an essential pillar of Astroparticle Physics that aims, in a unique way, to address many fundamental questions of the non-thermal Universe in the Astroparticle Physics domain. The huge physics potential of this field can be achieved by an upgrade of the performances of current ground-based experiments and pioneer space-based missions, as the JEM-EUSO space telescope \cite{jem-euso}. The JEM-EUSO space mission is the Extreme-Universe Space Observatory (EUSO) which will be located at the Exposure Facility of the Japanese Experiment Module (JEM/EF) on the International Space Station (ISS) and looking downward the atmosphere will allow a full-sky monitoring capability to watch for Ultra-High Energy Cosmic Rays (UHECR) and Extremely High Energy Cosmic Rays (EHECR). An Atmospheric Monitoring System (AMS) is mandatory and a key element of a Space-based mission which aims to detect Ultra-High Energy Cosmic Rays (UHECR). JEM-EUSO has a dedicated atmospheric monitoring system that plays a fundamental role in our understanding of the atmospheric conditions in the FoV of the main telescope. The JEM-EUSO AMS consists of a bi-spectral infrared camera and a LIDAR device that are being fully designed under space qualification to fulfil the scientific requirements of this space mission.  

\section{The Atmospheric Monitoring system}
\label{sec-AMS}

To fully monitor the atmosphere and to retrieve the cloud coverage and cloud top height in the JEM-EUSO FoV, an Atmospheric Monitoring System (AMS) is foreseen in the telescope. The AMS \cite{AMS} is crucial to estimate the effective UHECR \& EHECR exposure of the telescope and for the proper analysis of the UHECR \& EHECR events under cloudy conditions \cite{clouds}, \cite{clouds-2}.

The AMS of JEM-EUSO will include: 
\begin{enumerate}
\item
a bi-spectral Infrared (IR) camera;
\item
a LIght Detection And Ranging (LIDAR) device;
\item 
global atmospheric models generated from the analysis of all available meteorological data by global weather 
services such as the National Centers for Environmental Predictions (NCEP), the Global Modeling and 
Assimilation Office (GMAO) and the European Centre for Medium-Range Weather Forecasts (ECMWF)\cite{all}.
\item the slow mode data of JEM-EUSO, the monitoring of the pixel signal rate every 3.5 s for the observation 
of Transient Luminous Events (TLEs), which will give additional information on cloud distribution and the 
intensity of the night sky airglow.
\end{enumerate}
\noindent
The JEM-EUSO telescope will observe the EAS development only during night time. 
The IR camera will cover the entire FoV of the telescope in order to detect the presence 
of clouds and to obtain the cloud cover and the cloud top altitude. The LIDAR will be shot 
in some pre-defined position around the location of triggered EAS events. The LIDAR will be 
used to measure the clouds altitude and optical depth as well as the optical depth vertical 
profile of the atmosphere along these directions with a range accuracy of 375 m in nadir. The 
IR camera and the LIDAR have been designed to work in a complementary way.

\section{The Infrared Camera}
\label{sec-IRcamera}

\begin{figure}[h]
  \centering
  \includegraphics[width=0.5\textwidth]{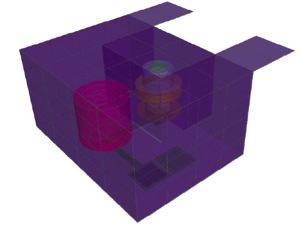}
  \caption{IR Camera Preliminary Design Modelization}
  \label{fig:imagen1}
 \end{figure}

The IR Camera (Fig.~\ref{fig:imagen1}) onboard JEM-EUSO will consist of a refractive optics made of germanium and an uncooled $\mu$bolometer array detector. The FoV of the IR Camera is $48^\circ$, totally matching the FoV of the main JEM-EUSO telescope. The angular resolution, which corresponds to one pixel, is about $0.1^\circ$. A temperature-controlled shutter in the camera and blackbodies are used to calibrate background noise and gains of the detector to achieve an absolute temperature accuracy of $\sim 3$ K. Therefore, the IR Camera takes images continuously every 17 s while the ISS moves 1/4 of the FoV of the JEM-EUSO telescope.

A brightness temperature measurement, intended for a single band infrared camera configuration, may not provide the required radiometric accuracy without the use of external information for atmospheric effects correction. Therefore a multispectral approach has been selected as baseline of the Infrared Camera of JEM-EUSO \cite{ICRC2013}. The delailed and dedicated retrieval algorithm for the cloud top height parameter that fulfill the scientific and technical requierements of the Infrared Camera of JEM-EUSO can be found in these Proceedings \cite{clouds-2}. In the preliminary design of the IR-Camera a bi-spectral design with two filters in the cold spot of the optics that allows a multispectral snapshot camera without a dedicated filter wheel mechanism is foreseen. This is a very “smart” solution that leads to have a more reliable baseline and to reduce the costs of a complicated filter wheel mechanism intended for Space applications. The only drawback of this solution is that, in order to overcome the use of half of available area of the detector for each spectral band, the IR-Camera images acquisition time has to be faster to avoid gaps during the ISS orbit with the impact on the restricted data budget allocated for an ISS mission.

\begin{figure}[h]
  \centering
  \includegraphics[width=0.4\textwidth]{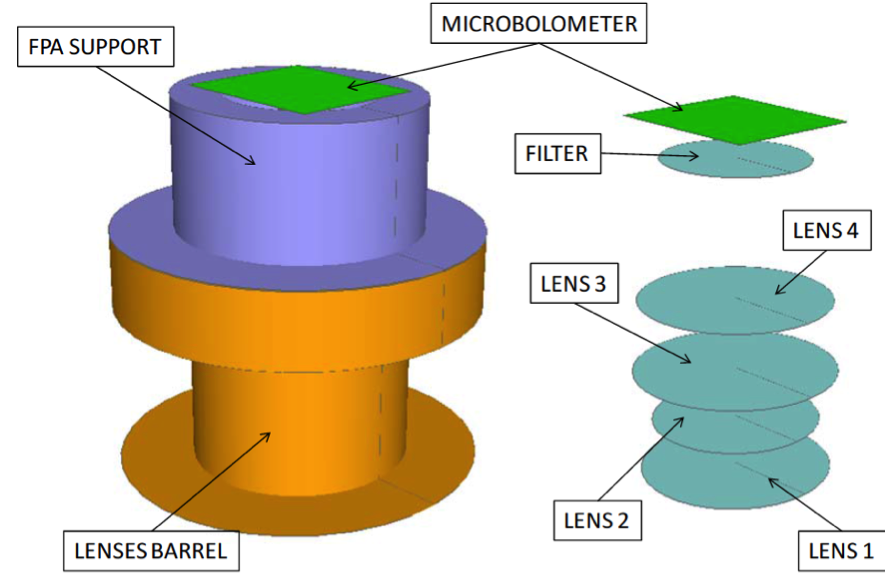}
  \caption{Preliminary Design Scheme of the Optical Unit Assembly.}
  \label{fig:optical}
 \end{figure}

\begin{figure}[h]
  \centering
  \includegraphics[width=0.4\textwidth]{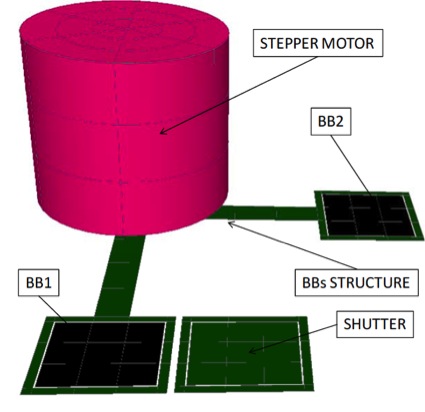}
  \caption{Scheme of the Calibration unit Assembly preliminary design with the two Black Bodies BB1 \& BB2 and the stepper motor}
  \label{fig:calibration}
 \end{figure}

\begin{figure}[h]
  \centering
  \includegraphics[width=0.5\textwidth]{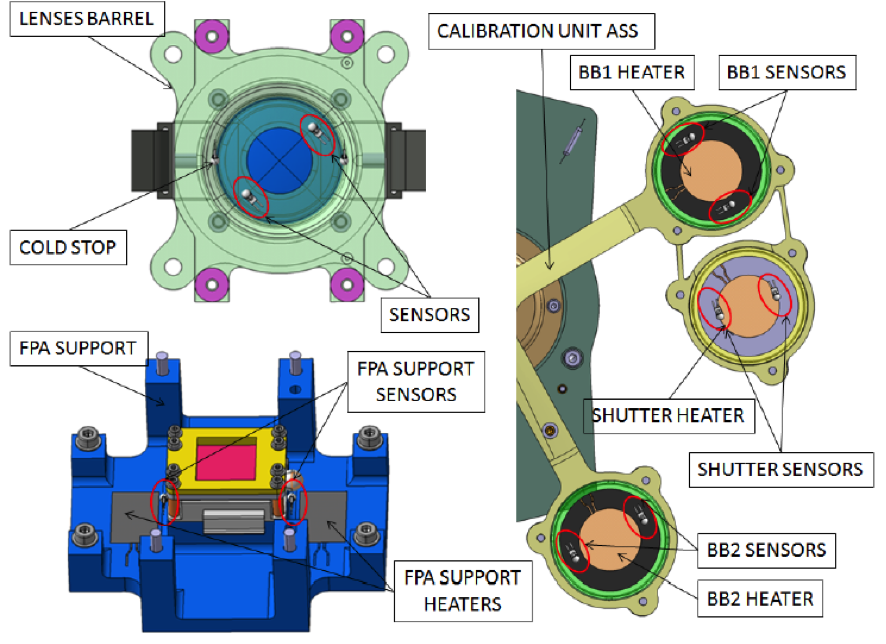}
\vspace{-0.4cm}
  \caption{Heaters \& Temperature Sensors Distribution in the System Preliminary Design of the Infrared Camera of JEM-EUSO.}
  \label{fig:heaters}
\vspace{-0.4cm}
 \end{figure}

The preliminary design of the IR Camera \cite{ICRC2013} can be divided into three main blocks: the Telescope Assembly, the Calibration Unit and the Electronic Assembly. The main function of the Telescope Assembly is to acquire the infrared radiation by means of an uncooled microbolometer and to convert it into digital counts. A dedicated optical design has been developed as well, with a huge angular field to comply with the wide FoV of the JEM-EUSO main telescope (Fig.~\ref{fig:optical}). To assure the high demanding accuracy, a dedicated on-board calibration system is foreseen (Fig.~\ref{fig:calibration}). Moreover, this System Preliminary Design is complemented by a challenging Mechanical and Thermal design to secure that the IR-Camera will be completely isolated (Fig.~\ref{fig:heaters}).

\begin{figure}[h]
  \centering
  \includegraphics[width=0.4\textwidth]{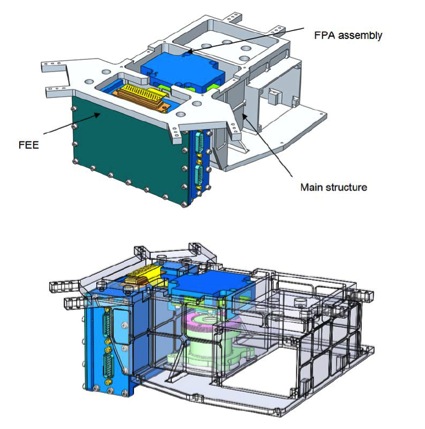}
  \caption{FEE overview of the Infrared Camera.}
  \label{fig:FEE}
 \end{figure}

Meanwhile, the Electronic Assembly provides mechanisms to process and transmit the obtained images, the electrical system, the thermal control and to secure the communication with the platform computer. The Electronics Assembly will be communicated with the Main Instrument (apart from the power input buses) by means of three main links: a Data link, a Command link, and Synchronization lines. In addition, the Electronics Assembly will provide discrete telemetries to the Main Instrument. Regarding the Front End Electronics (FEE), they will be also communicated through serial links, synchronization lines and power generated in the Electronics Assembly to supply the FEE (Fig.~\ref{fig:FEE}).

Moreover, the Interface Control Unit (ICU) can be separated into 2 blocks: the controlling board block and the driving board block. The controlling board block is composed of the electronics board in which will run the communication protocol, using a RS422 type as “telemetry and command" bus. This controlling board is also in charge of the compression algorithm used for the images previously captured by the optics to be sent to the Main Instrument. Assessment of the VHDL code occupation has been made giving as a result that at least a RTAX2000S FPGA from Actel is needed to implement the compression algorithm and the communications protocol. The Driving Board will be composed of the board which contains the driver for the motor and the switches acquisition. Both functionalities will be managed by a small FPGA which runs the control algorithm and drives the actuator. Occupation assessment has been performed, giving as a result that the RTSX72-SU FPGA from Actel is suitable for this design. In addition, a motor will be connected to drivers through relays. These relays will isolate the motor from electronics when stands off. Otherwise, some currents from non-active winding could be driven to the non-active electronics. Driving board will provide power to the heaters (32.2 $\Omega$ resistance with a nominal power dissipation of 1.69 W) placed in calibration unit by means of a 1 Hz Pulse Width Modulation (PWM) line of 15 V.

\begin{figure}[h]
  \centering
  \includegraphics[width=0.5\textwidth]{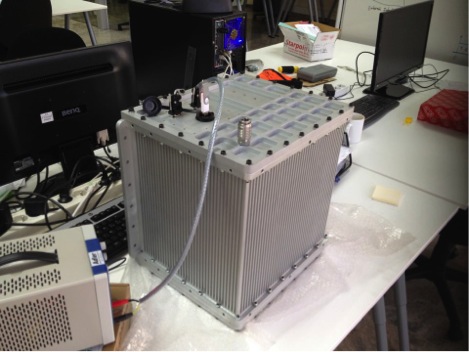}
  \caption{Dedicated bi-spectral and waterproof Infrared Camera for the EUSO-BALLOON (CNES) pathfinder.}
  \label{fig:camara-balloon}
 \end{figure}

A full characterization and calibration of the Bread Board Model (BBM) for the Infrared Camera of the EUSO-Balloon pathfinder led by CNES \cite{balloon} was performed on April 2014 at the Instituto de Astrofisica de Canarias (IAC, Tenerife). Moreover this dedicated bi-spectral and waterproof Infrared Camera was flown with EUSO-BALLOON first flight on August 24, 2014 from Timmins (Canada) (Fig.~\ref{fig:camara-balloon}). 

\section{The LIDAR}
\label{sec-lidar}
The task of the LIDAR is to localize optically thin clouds and aerosol layers and to provide 
measurements of the scattering and absortion properties of the atmosphere in the region 
of the EAS development and between the EAS and the JEM-EUSO telescope. 

\subsection{LIDAR design}
\label{subsec-lidar-hardware}
The LIDAR is composed of a transmission and a receiving system. The transmission system comprises 
a Nd:YAG laser and a pointing mechanism (PM) to steer the laser beam in the direction of the triggered EAS events. 
As the laser backscattered signal will be received back by the JEM-EUSO telescope (working as the LIDAR receiver), 
the laser operational wavelength was chosen to be the third harmonic of the Nd:YAG laser, at $\lambda = $~355~nm. 
The laser is being developed at RIKEN (Japan) and will be part of the JAXA (Japanese Space Agency) contribution to the Mission. The 
PM is under development at the University of Geneva, in close collaboration with CSEM\footnote{www.csem.ch} 
(Switzerland).
 
\begin{figure}[h]
\centering
\includegraphics[width=0.45\textwidth]{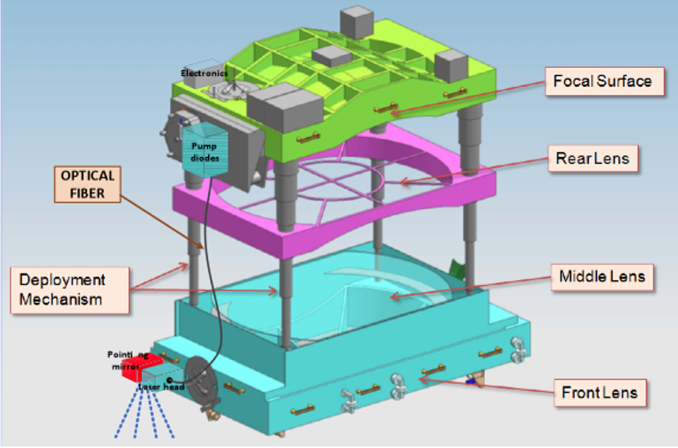}
\caption{Schematic LIDAR placement on the JEM-EUSO telescope. The laser head and the pointing mirror will be 
placed close to the front lens of the telescope, while the pump diodes and the control electronics for the mirrors 
will be placed on the focal surface.}
\label{fig:LIDARonJEM-EUSO}    
\end{figure}

In the current JEM-EUSO design, the light of the pumping diodes is guided to the laser head through an optical fibre. 
The PM is conceived to have a steering mirror with two angular degrees of freedom and a maximal tilting angle 
of $\pm$~15$^\circ$, needed to move the laser beam anywhere within the JEM-EUSO FoV. The LIDAR system will be 
integrated into the JEM-EUSO telescope. A preliminary scheme of the placement of the different elements 
is shown in Fig.~\ref{fig:LIDARonJEM-EUSO}. A summary of the specifications needed for the entire system 
is reported in Tab.~\ref{tab:LIDAR}. 

\begin{figure}[h]
\centering
\includegraphics[width=7cm]{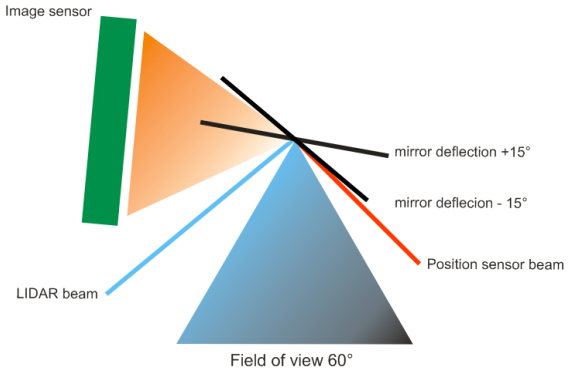}
\caption{Representation of the laser and position control beam, facing the front side of the MEMS mirror.}
\label{fig:mirror2}    
\end{figure}

\begin{figure}[h]
\centering
\includegraphics[width=8.2cm]{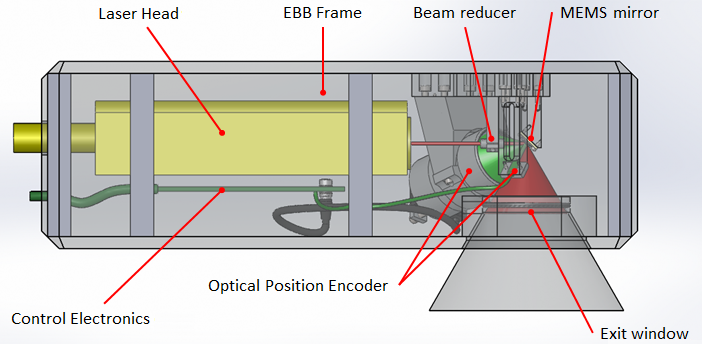}
\caption{CAD model of the LIDAR PM EBB. In the figure we show the housing, the commercial version of the laser that will be used for the 
EBB, the optical system supporting the positional sensor, the mems mirror, and the electronic board. All components are represented 
in scale (the MEMS mirror is about 3$\times$3~mm large). The red beam represents the laser beam, while the green one is the beam used 
to measure the position of the MEMS mirror through the positional sensor and the supporting optical system.}
\label{fig:EBB}    
\end{figure}

The LIDAR is expected to receive on average a new trigger on possible EAS events roughly every $\sim$10 s. 
In the time between two consecutive triggers, the PM should be able to decode the information on the location 
of the last triggered event within the telescope FoV, re-point the laser beam in this direction, and shoot  
5 laser shots covering a sufficiently wide region around the EAS position. The effective time available to the pointing system 
to steer the laser beam is thus typically of few tenths of seconds, thus requiring a lightweight mirror with limited inertia 
to optimize the operations of the PM. For this reason, the innovative Micro-Electro-Mechanical Systems (MEMS) technology 
has been selected by the UniGe to develop the tilting mirror \citep{bayat2012}. 
\begin{figure}[h]
\centering
\includegraphics[width=7cm]{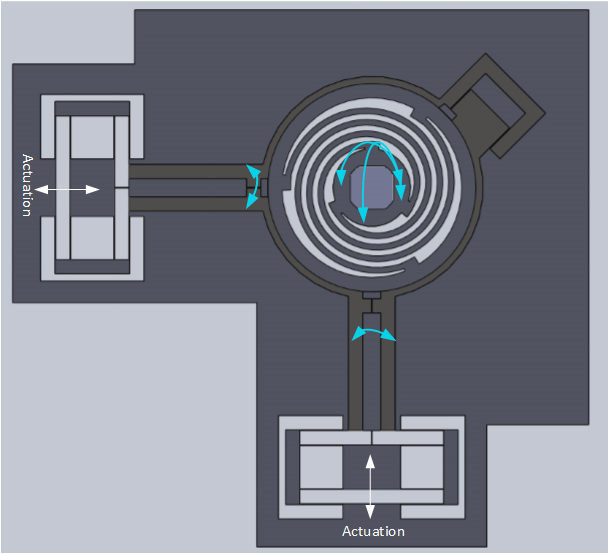}
\caption{CAD model of the MEMS mirror (not in scale, view from the bottom side). The mirror is squezed between two support structures.  
It is hold through silicon beamers that allow tilting movements when the levers located on the bottom supporting structure 
are forced in the X and Y direction by the magnetic forces generated by two magnets and the coils controlled by a dedicated 
electronic board. The coils are not sketched. The two magnets are not visible, but located at the bottom of the actuation points.}
\label{fig:mirror}    
\end{figure} 
This mechanism will use magnetic forces to achieve the steering pointing function. 
A CAD design of the device has been already prepared and is presented in Fig.~\ref{fig:mirror}. 
It consists of a sandwich of two supporting structures that allow a tip-tilt movement of the mirror squeezed in between. 
The lower supporting structure can be forced in the X and Y directions independently by levers connected to it. The
ends of these levers are connected to guided magnets that are actuated in the X and Y directions by coils. 
The coils will be controlled through an electronic control board, that is also in charge to receive the trigger 
for possible EAS events and translate this into mirror deplacements. The mirror position is controlled at any time 
through a high precision positional sensor and the auxiliary optical system. Both the laser and the beam used for the positional sensor 
will be facing the front side of the mirror, as sketched in Fig.~\ref{fig:mirror2}. 
The development of the PM at the UniGe is currently in phase B1. This phase is expected to be completd at the end 
of 2014 and encompasses the complete design of an Elegant BreadBoard (EBB, see Fig.~\ref{fig:EBB}), whose scope is to demonstrate 
the achievement of the critical PM functionalities. The EBB comprises a commercial version of the  
laser operating at the same wavelenght of the JEM-EUSO RIKEN laser, an optical system with a position encoder, the MEMS mirror prototype, 
and the control electronics. The EBB will be implemented during phase B2 and it is expected that a fully operational model  
will be available at the end of 2015. A qualification and flight model of the JEM-EUSO PM are planned for $>$2016 (project phase C/D).  

\begin{table}[h]
\centering
\caption{Specification for the JEM-EUSO LIDAR.}
\label{tab:LIDAR}      
\begin{tabular}{ll}
\hline\noalign{\smallskip}
Parameter			&		 Specification				\\
\noalign{\smallskip}\hline\noalign{\smallskip}
Wavelength 			& 		$355$ nm				 	\\
Repetition Rate	 	& 		$1$ Hz					 \\
Pulse width   	   		& 		$15$ ns		            		\\
Pulse energy 	  		& 		$20$ mJ/pulse		     		\\
Beam divergence		&		 $0.2$ mrad 				\\
Receiver        		 	& 		JEM-EUSO telescope	 	\\
Detector                 	 	& 		MAPMT (JEM-EUSO)	  	\\
Range resolution (nadir) 	&		 $375$ m 	   			  	\\
Steering of output beam 	&		 $\pm 30^\circ$ from vertical 	\\
Mass   				&		 $14$ kg				     	\\
Dimension  			&		 $450\times350\times250$ mm\\
Power				&		$< 20$ W					\\
\noalign{\smallskip}\hline
\end{tabular}
\end{table}

\subsection{Simulations and data analysis}

\begin{figure}[h]
  \centering
  \includegraphics[width=0.5\textwidth]{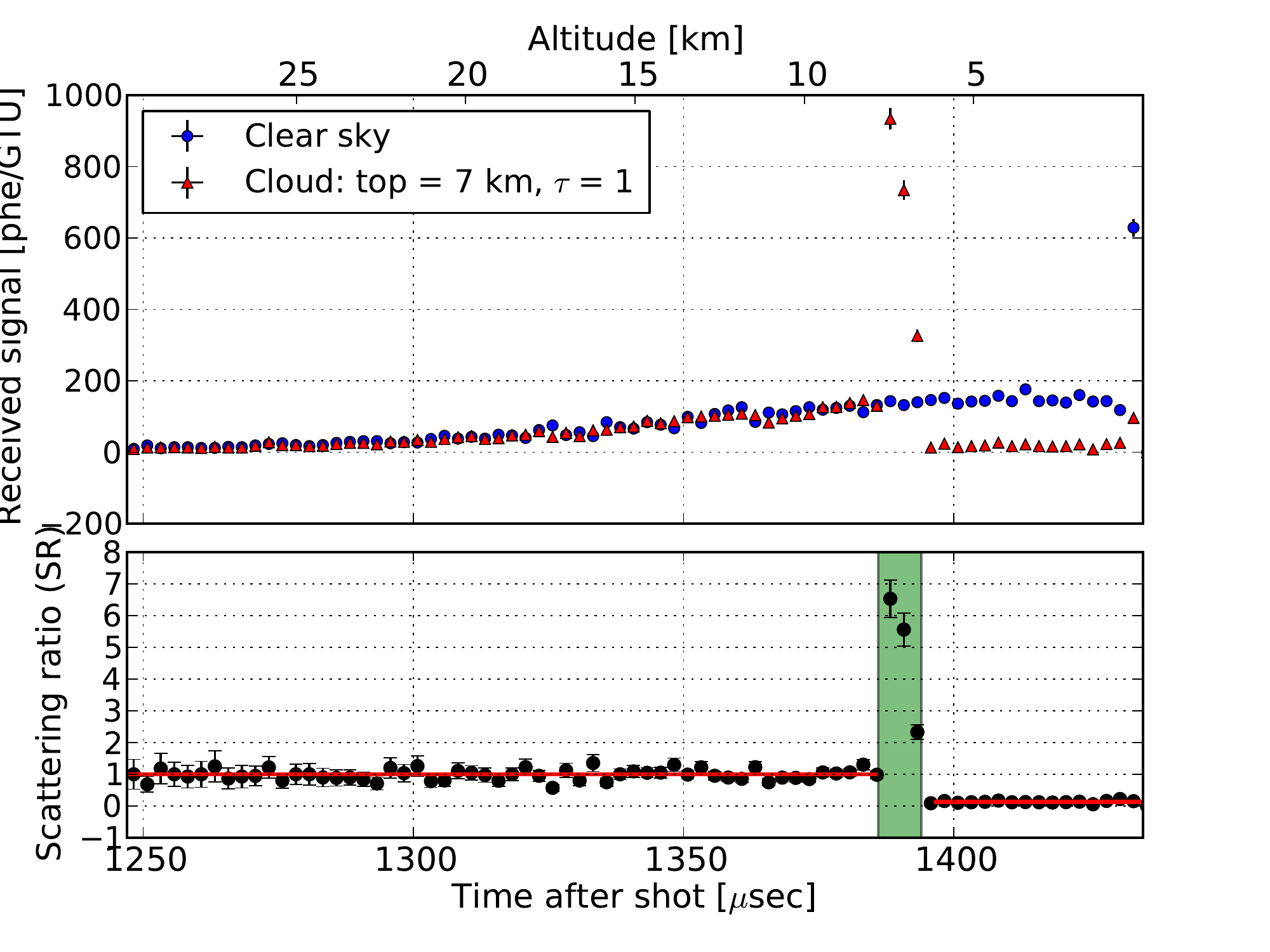}
  \caption{\emph{Top}: LIDAR backscattered signal in clear sky (blue) and in the presence of an optically 
  thick ($\tau = 1$) cloud (red) as a function of time. \emph{Bottom}: Scattering ratio (SR).}
  \label{fig:LIDAR}
 \end{figure}

Simulations \cite{Toscano_ICRC2013} have been carried out in order to study the capability of the system in retrieving the 
physical properties of atmospheric features such as clouds or aerosol layers. Fig.~\ref{fig:LIDAR} 
shows an example of the simulated laser backscattered signal detected by the JEM-EUSO telescope. 

\label{subsec-analysis}
The top panel shows the signal in case of clear atmosphere (blue circles) and in presence of 
the cloud (red triangles) as a function of the time after shooting the laser and the altitude. 
The presence of a cloud at $\sim7$~km will be clearly detected by the LIDAR as an increase of 
the backscattered signal coming from that region. The bottom panel shows the so-called LIDAR 
Scattering Ratio (SR), the ratio between the backscattered signal detected in the real condition 
and a reference profile which represents the backscattered signal in clear atmosphere.
Fitting the SR in the region below the cloud allows for the measurement of the optical depth 
simply using the formula $\tau = -log(SR) / 2$. 

\begin{figure}[h]
  \centering
  \includegraphics[width=0.5\textwidth]{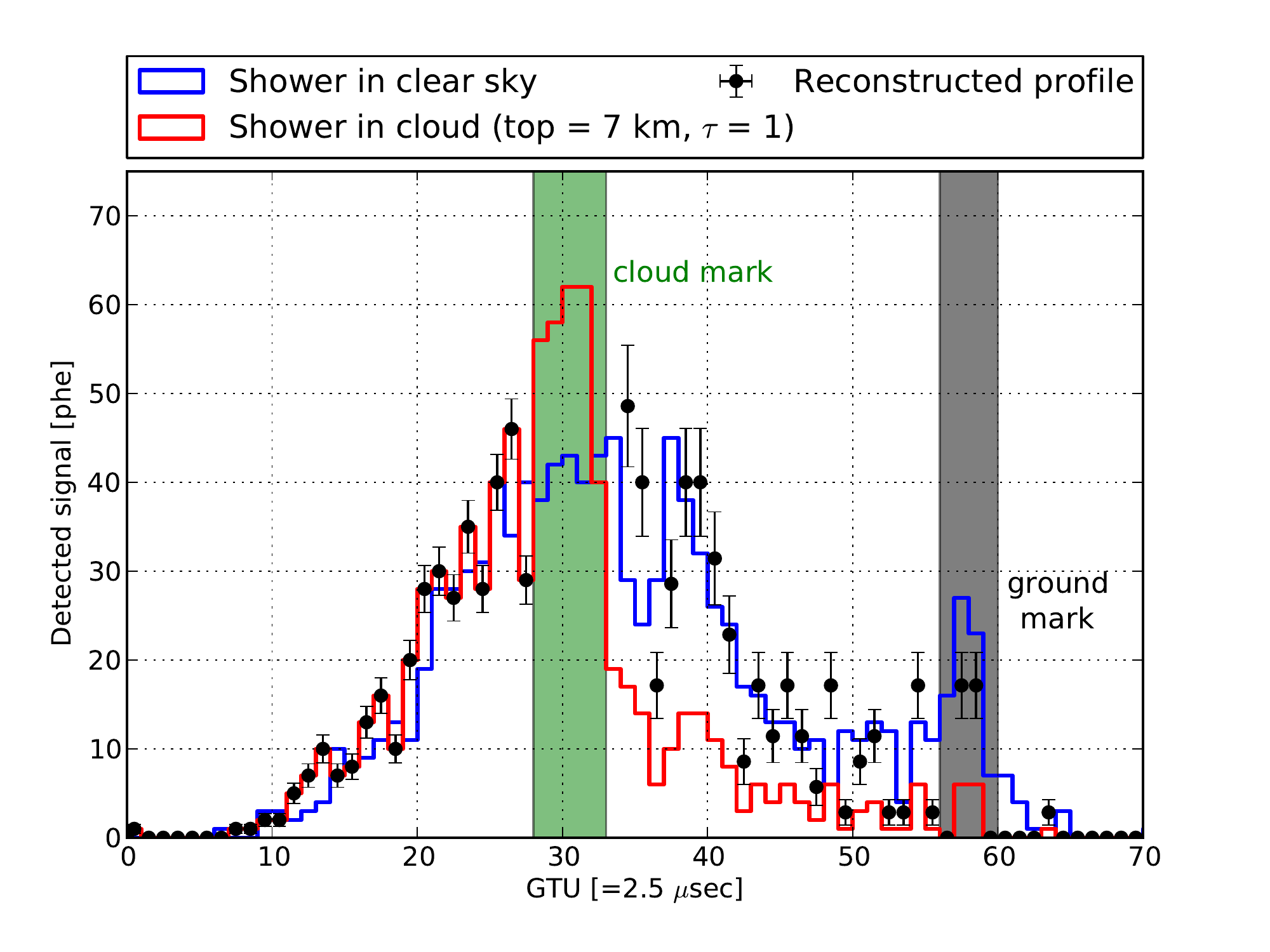}
  \caption{Reconstructed time profile (black points) of $10^{20}$ eV EAS together with the clear 
  atmosphere (blue) and cloud affected (red) profiles. Error bars are statistical only.}
  \label{fig:RecoProfile}
 \end{figure}

Once the cloud is detected and its optical depth determined, the cloud-affected EAS profile 
can be corrected using the formula: 
\begin{equation}
\label{eq:Shower}
Signal_{cloud} = Signal_{clear} (\exp^{-\tau}). 
\end{equation}
Fig.~\ref{fig:RecoProfile} shows the photons arrival time in GTU\footnote{The Gate Time Unit, or 
GTU, is the time unit of the detector focal surface; 1~GTU corresponds to 2.5~$\mu$sec.} at the 
detector focal plane for a shower generated by a UHE proton with $E = 10^{20}$~eV and 
$\theta = 60^\circ$. 
The blue histogram represents the profile of the shower developing in a clear atmosphere, 
characterised by the presence of the ``ground mark" at $\sim$60~GTU. This feature is due to 
Cherenkov photons hitting the ground and reflected back to the JEM-EUSO focal surface. The 
red histogram shows the profile of the shower crossing the same optically thick cloud shot 
by the LIDAR and located at an altitude of $\sim$7~km. As in the case of the LIDAR, the 
presence of the cloud modifies the EAS time profile, with the appearance of a new feature 
(the ``cloud mark") at $\sim$~28~GTU, and the ground mark vanishing. After the correction 
is done using the LIDAR measurement it is possible to retrieve the correct profile (black points) 
and almost entirely recover the ground mark feature.     

\section{Conclusions}
\label{sec-conclusions}

The Infrared Camera and LIDAR of the JEM-EUSO Space Mission are under fully design, prototyping and development under responsability of Japan, Switzerland and Spain. Presently both devices are under Preliminary Design Phase and space qualification of the Infrared Camera and the LIDAR are foreseen to accomplish for the scientific and technical specifications of the JEM-EUSO Space Mission.
 
\section*{Acknowledgements}

The JEM-EUSO team at the University of Geneva acknowledges support from the Swiss Space Office through a dedicated PRODEX program. This work is supported by the Spanish Government MICINN \& MINECO under the Space Program: projects AYA2009-06037-E/AYA, AYA-ESP 2010-19082, AYA-ESP 2011-29489-C03, AYA-ESP 2012-39115-C03, AYA-ESP 2013-47816-C4, MINECO/FEDER-UNAH13-4E-2741, CSD2009-00064 (Consolider MULTIDARK) and by Comunidad de Madrid under projects S2009/ESP-1496 \& S2013/ICE-2822. M. D. Rodriguez Frias acknowledge International Visitor Grant from the Swiss National Science Foundation (SNSF). 


\begin{thebibliography}{}
%

\bibitem{jem-euso}
Adams Jr., J.H. et al. (The JEM-EUSO Collaboration), 
An evaluation of the exposure in nadir observation of the JEM-EUSO mission.
\textbf Astroparticle Physics, 44, 76, 90 (2013). 

Rodr\'iguez Fr\'ias, M. D. et al. for the JEM-EUSO Collaboration,
The JEM-EUSO Space Mission: Frontier Astroparticle Physics @ ZeV range from Space.
\textbf  Homage to the Discovery of Cosmic Rays. Nova Science Publishers, New York, ISBN: 978-1-62618-998-0, Inc, Pg 201-212 (2013).

\bibitem{AMS}
Rodr\'iguez Fr\'ias, M. D. et al. for the JEM-EUSO Collaboration. 
The Atmospheric Monitoring System of the JEM-EUSO Space Mission.
\textbf Proc. International Symposium on Future Directions in UHECR Physics, The European Physical Journal, Vol 53, 10005-pg1-7,  http://dx.doi.org/10.1051/epjconf/20135310005, (2013).

The JEM-EUSO Collaboration (corresponding authors S. Toscano, J. A. Morales de los Rios, A. Neronov, M. D. Rodr\'iguez Fr\'ias \& S. Wada). The Atmospheric Monitoring System of the JEM-EUSO instrument.
\textbf Experimental Astronomy {37}, DOI 10.1007/s10686-014-9378-1 (2014).

\bibitem{clouds}
S\'aez-Cano, G., Shinozaki, K., del Peral, L., Bertaina, M. and Rodr\'iguez Fr\'ias, M.D. for the JEM-EUSO Collaboration. Observation of extensive air showers in cloudy conditions by the JEM-EUSO
Space Mission. 
\textbf Advance in Space Research, {53}, 1536-1543 (2014).  

\bibitem{clouds-2}
S\'aez-Cano, G., Morales de los Rios, J. A., del Peral, L., Neronov, A., Wada, S. and Rodr\'iguez Fr\'ias, M.D. for the JEM-EUSO Collaboration. Thin and thick cloud top height retrieval algorithm with the Infrared Camera and LIDAR of JEM-EUSO.
\textbf These Procc.
 
\bibitem{all}
http://www.ncep.noaa.gov/; http://gmao.gsfc.nasa.gov/; http://www.ecmwf.int/

\bibitem{ICRC2013}
Rodr\'iguez Fr\'ias, M. D. et al. for the JEM-EUSO Collaboration. 
Towards the Preliminary Design Review of the Infrared Camera of the JEM-EUSO Space Mission.
\textbf Proc. of 33rd International Cosmic Ray Conference (ICRC), Rio de Janeiro, Brazil (2013), arXiv:1307.7071v1 [astro-ph.IM]  

The JEM-EUSO Collaboration (corresponding authors J. A. Morales de los Rios \& M. D. Rodr\'iguez Fr\'ias). The Atmospheric Monitoring System of the JEM-EUSO instrument.
\textbf Experimental Astronomy {37}, DOI 10.1007/s10686-014-9402-5 (2014).

\bibitem{balloon}
Von Ballmoos, P., et al. A balloon-borne prototype for demonstrating the concept of JEM-EUSO.
\textbf Advance in Space Research {53}	, 1544-1560 (2014).                                    

Morales de los Ríos, J. A., Joven, E., del Peral, L., Reyes, M., Licandro, J., and Rodríguez Frías, M. D.. 
The Infrared Camera Prototype Characterization for the JEM-EUSO Space Mission.
\textbf Nuclear Instruments and Methods NIMA, {749}, 74-83,  ISSN 0168-9002 (2014). 

\bibitem{bayat2012}
D. Bayat, ``Large Hybrid High Precision MEMS Mirrors'', PhD Thesis, 2012 
(http://infoscience.epfl.ch/record/167903/files/ \\ EPFL\_TH5152.pdf)

\bibitem{Toscano_ICRC2013} 
S. Toscano, L. Valore, A. Neronov, F. Guarino (JEM-EUSO Collaboration), ``LIDAR treatment inside the ESAF. ''Simulation Framework for the JEM-EUSO mission'',  
\textbf Proc. of 33rd International Cosmic Ray Conference (ICRC), Rio de Janeiro, Brazil (2013), ID0530.
Preprint: arXiv:1307.7071.
\end{thebibliography}
\end{document}